\documentclass[journal]{IEEEtran}
\usepackage{cite}
\usepackage{enumerate}
\usepackage{graphics}
\usepackage{caption3}
\usepackage{commath}
\usepackage[dvipsnames]{xcolor}
\usepackage{comment}
\usepackage{amsmath,amssymb,amsfonts}
\usepackage{algorithmic}
\usepackage{graphicx}
\usepackage{mathtools}
\usepackage{siunitx}
\usepackage{booktabs,array}
\usepackage{epsfig}
\usepackage{epstopdf}
\usepackage{listings}
\usepackage[caption=false]{subfig}
\usepackage{soul}
\usepackage{textcomp}
\def\BibTeX{{\rm B\kern-.05em{\sc i\kern-.025em b}\kern-.08em
    T\kern-.1667em\lower.7ex\hbox{E}\kern-.125emX}}

\DeclareMathOperator*{\argmin}{arg\,min}

\newcommand{\smallsection}[1]{\noindent\textbf{#1}}

\usepackage[acronym,shortcuts,nonumberlist,nopostdot]{glossaries}
\usepackage{glossary-mcols}

\makenoidxglossaries

\newacronym{4g}{4G}{fourth-generation}
\newacronym{5g}{5G}{fifth-generation}
\newacronym{3gpp}{3GPP}{Third-generation partnership project}
\newacronym{abft}{A-BFT}{association beamforming training}
\newacronym{abt}{ABT}{asymmetric beamforming training}
\newacronym{ad}{AD}{angle-Doppler}
\newacronym{adc}{ADC}{analog-to-digital converter}
\newacronym[firstplural=angle differences-of-arrival]{adoa}{ADoA}{angle difference-of-arrival}
\newacronym{ae}{AE}{autoencoder}
\newacronym[firstplural=angles of arrival]{aoa}{AoA}{angle of arrival}
\newacronym[firstplural=angles of departure]{aod}{AoD}{angle of departure}
\newacronym{ap}{AP}{access point}
\newacronym{api}{API}{application program interface}
\newacronym{ar}{AR}{augmented reality}
\newacronym{bi}{BI}{beacon interval}
\newacronym{brp}{BRP}{beam refinement protocol}
\newacronym{bs}{BS}{base station}
\newacronym{bti}{BTI}{beacon transmission interval}
\newacronym{cacfar}{CA-CFAR}{cell-averaging constant false alarm rate}
\newacronym{cbap}{CBAP}{contention based access period}
\newacronym{cdf}{CDF}{cumulative distribution function}
\newacronym{cir}{CIR}{channel impulse response}
\newacronym{cnn}{CNN}{convolutional NN}
\newacronym{com}{COM}{center of mass}
\newacronym{cots}{COTS}{commercial off-the-shelf}
\newacronym{csi}{CSI}{channel state information}
\newacronym{dac}{DAC}{digital-to-analog converter}
\newacronym{dbscan}{DBSCAN}{density-based spatial clustering of applications with noise}
\newacronym{dft}{DFT}{discrete Fourier transform}
\newacronym{dkf}{DKF}{discrete KF}
\newacronym{dl}{DL}{deep learning}
\newacronym{dsp}{DSP}{digital signal processor}
\newacronym{dti}{DTI}{data transmission interval}
\newacronym{ecd}{ECD}{expand-contract dilation}
\newacronym{ekf}{EKF}{extended Kalman filter}
\newacronym{em}{EM}{expectation maximization}
\newacronym{endc}{EN-DC}{enhanced UTRA-dual connectivity}
\newacronym{esprit}{ESPRIT}{estimation of signal parameters via rotational invariance techniques}
\newacronym{fcos}{FCOS}{fully connected \mbox{one-stage}}
\newacronym{fft}{FFT}{fast Fourier transform}
\newacronym{fmcw}{FMCW}{frequency-modulated continuous wave}
\newacronym{fpga}{FPGA}{field-programmable gate array}
\newacronym{fscn}{FSCN}{fractionally strided convolutional network}
\newacronym{ftm}{FTM}{fine time measurement}
\newacronym{gan}{GAN}{generative adversarial network}
\newacronym{gru}{GRU}{gated recurrent unit}
\newacronym{gscm}{GSCM}{geometry-based stochastic channel model}
\newacronym{hdc}{HDC}{hybrid dilated convolution}
\newacronym{hr}{HR}{heart rate}
\newacronym{hvrae}{HVRAE}{hybrid variational RNN autoencoder}
\newacronym{if}{IF}{intermediate-frequency}
\newacronym{ifft}{IFFT}{inverse FFT}
\newacronym{iir}{IIR}{infinite impulse response}
\newacronym{iot}{IoT}{Internet of things}
\newacronym{itur}{ITU-R}{International telecommunication union -- radiocommunication Sector}
\newacronym{kf}{KF}{Kalman filter}
\newacronym{lm}{LM}{Levenberg-Marquardt}
\newacronym{lms}{LMS}{least mean squares}
\newacronym{los}{LoS}{line-of-sight}
\newacronym{lstm}{LSTM}{long-short term memory}
\newacronym{mac}{MAC}{medium access control}
\newacronym{mcu}{MCU}{micro-controller unit}
\newacronym{mf}{MF}{matched filter}
\newacronym{mimo}{MIMO}{multiple-input multiple-output}
\newacronym{miso}{MISO}{multiple-input single-output}
\newacronym{ml}{ML}{machine learning}
\newacronym{mmse}{MMSE}{minimum mean-square error}
\newacronym{mmw}{mmWave}{millimeter-wave}
\newacronym{mpc}{MPC}{multipath component}
\newacronym{mse}{MSE}{mean-square error}
\newacronym{music}{MUSIC}{multiple signal classification}
\newacronym{nis}{NIS}{normalized innovation squared}
\newacronym{nlos}{NLoS}{non-line-of-sight}
\newacronym{nn}{NN}{neural network}
\newacronym{noma}{NOMA}{non-orthogonal multiple access}
\newacronym{ofdm}{OFDM}{orthogonal frequency-division multiplexing}
\newacronym{oks}{OKS}{object keypoints similarity}
\newacronym{pa}{PA}{power amplifier}
\newacronym{pdp}{PDP}{power-delay profile}
\newacronym{phy}{PHY}{physical layer}
\newacronym{prs}{PRS}{positioning reference signal}
\newacronym{psd}{PSD}{power spectral density}
\newacronym{pw}{PW}{pulsed wave}
\newacronym{ra}{RA}{range-azimuth}
\newacronym{rd}{RD}{range-Doppler}
\newacronym{rda}{RDA}{range-Doppler-azimuth}
\newacronym{relu}{ReLU}{rectified linear unit}
\newacronym{resnet}{ResNet}{residual network}
\newacronym{rf}{RF}{radio frequency}
\newacronym{rgb}{RGB}{red-green-blue}
\newacronym{rms}{RMS}{root mean square}
\newacronym{rmse}{RMSE}{root mean square error}
\newacronym{rnn}{RNN}{recursive NN}
\newacronym{rss}{RSS}{received signal strength}
\newacronym{rssi}{RSSI}{received signal strength indicator}
\newacronym{rtt}{RTT}{round-trip time}
\newacronym{rx}{RX}{receiver}
\newacronym{rzf}{RZF}{regularized zero-forcing}
\newacronym{saf}{SAF}{spatial attention fusion}
\newacronym{sage}{SAGE}{space-alternating generalized expectation maximization}
\newacronym{sdr}{SDR}{software-defined radio}
\newacronym{slam}{SLAM}{simultaneous localization and mapping}
\newacronym{sls}{SLS}{sector-level sweep}
\newacronym{snr}{SNR}{signal-to-noise ratio}
\newacronym{soc}{SoC}{system-on-chip}
\newacronym{sp}{SP}{service period}
\newacronym{spi}{SPI}{subsample peak interpolation}
\newacronym{srs}{SRS}{sounding reference signal}
\newacronym{ssw}{SSW}{sector sweep}
\newacronym{sta}{STA}{station}
\newacronym{stft}{STFT}{short-time Fourier transform}
\newacronym{sv}{SV}{Saleh-Valenzuela}
\newacronym{svm}{SVM}{support vector machine}
\newacronym{tdoa}{TDoA}{time difference-of-arrival}
\newacronym{toa}{ToA}{time of arrival}
\newacronym{tof}{ToF}{time of flight}
\newacronym{tx}{TX}{transmitter}
\newacronym{txss}{TXSS}{transmit sweep}
\newacronym{uav}{UAV}{unmanned aerial vehicle}
\newacronym{ue}{UE}{user equipment}
\newacronym{ukf}{UKF}{unscented Kalman filter}
\newacronym{urllc}{URLLC}{ultra-reliable low-latency communications}
\newacronym{usrp}{USRP}{universal software radio peripheral}
\newacronym{v2x}{V2X}{vehicle-to-everything}
\newacronym{va}{VA}{virtual anchor}
\newacronym{vco}{VCO}{voltage-controlled oscillator}
\newacronym{vr}{VR}{virtual reality}
\newacronym{wlan}{WLAN}{wireless LAN}
\newacronym{zf}{ZF}{zero-forcing}

\setglossarystyle{alttree}
\glssetwidest{URLL-NMP}

\ifCLASSINFOpdf
 
\else

\fi

\begin{document}
\bstctlcite{refLOC:BSTcontrol}

\title{Indoor Millimeter Wave Localization using Multiple Self-Supervised Tiny Neural Networks}

% \author{Anish~Shastri,~\IEEEmembership{Graduate Student Member,~IEEE,} Andres Garcia-Saavedra,~\IEEEmembership{Member,~IEEE,}
% Paolo~Casari,~\IEEEmembership{Senior~Member,~IEEE}% <-this % stops a space
\author{Anish~Shastri, Andres Garcia-Saavedra,
Paolo~Casari% <-this % stops a space
\thanks{This work received support from the European Commission's Horizon 2020 Framework Programme under the Marie Sk{\l}odowska-Curie Action MINTS (GA no.~861222).
}
\thanks{A. Shastri and P. Casari are with the Department
of Information Engineering and Computer Science, University of Trento, Italy. 
E-mails: \{anish.shastri, paolo.casari\}@unitn.it}
\thanks{A. Garcia-Saavedra is with NEC Laboratories Europe, Heidelberg, Germany. E-mail: andres.garcia.saavedra@neclab.eu}
% <-this % stops a space
% <-this % stops a space
}

% The paper headers
\markboth{Journal of \LaTeX\ Class Files,~Vol.~14, No.~8, August~2023}%
{Shell \MakeLowercase{\textit{et al.}}: Bare Demo of IEEEtran.cls for IEEE Journals}
% If you want to put a publisher's ID mark on the page you can do it like
% this:
%\IEEEpubid{0000--0000/00\$00.00~\copyright~2015 IEEE}
% Remember, if you use this you must call \IEEEpubidadjcol in the second
% column for its text to clear the IEEEpubid mark.

% use for special paper notices
%\IEEEspecialpapernotice{(Invited Paper)}

% make the title area
\maketitle

% As a general rule, do not put math, special symbols or citations
% in the abstract or keywords.
\begin{abstract}

We consider the localization of a mobile \acrlong{mmw} client in a large indoor environment using multilayer perceptron \acp{nn}. Instead of training and deploying a single deep model, we proceed by choosing among multiple tiny \acp{nn} trained in a self-supervised manner. The main challenge then becomes to determine and switch to the best \ac{nn} among the available ones, as an incorrect \ac{nn} will fail to localize the client. In order to upkeep the localization accuracy, we propose two switching schemes: one based on a Kalman filter, and one based on the statistical distribution of the training data. We analyze the proposed schemes via simulations, showing that our approach outperforms both geometric localization schemes and the use of a single \ac{nn}.
%, achieving sub-meter localization errors in \as{90\%} of the cases.

\end{abstract}

% Note that keywords are not normally used for peerreview papers.
% \begin{IEEEkeywords}
% mmWaves, self-supervised learning, Kalman filters, angle-difference of arrival, Mahalanobis distance.
% \end{IEEEkeywords}

% For peerreview papers, this IEEEtran command inserts a page break and
% creates the second title. It will be ignored for other modes.
\IEEEpeerreviewmaketitle

\section{Introduction}
\label{sec:intro}

\glsunset{mmw}

\IEEEPARstart{M}{illimeter} wave (\acrshort{mmw}) technologies support not just multi-Gbit/s data rates for next-generation applications such as \ac{ar} and 8-K video streaming, but also high-accuracy location systems~\cite{comst2023anish}. The use of large antenna arrays results in highly directional beams, leading to quasi-optical propagation. Because \ac{mmw} communications are short-ranged due to atmospheric attenuation (especially around the 60~GHz band), and are easily blocked by obstacles, it is common to consider dense deployments of \ac{mmw} \acp{ap}~\cite{mmwaveComst2017Hanzo}.  % especially in large indoor environments
In such scenarios, location information becomes an extremely useful tool to optimize the performance of \ac{mmw} networks~\cite{scalingmmWave2019Fiandrino}.

Existing localization algorithms employ the geometric properties of \ac{mmw} signals such as the \ac{aoa}, \ac{aod}, and \ac{tof} to localize a \ac{mmw} device. However, they require the knowledge of the indoor area, e.g., the locations of the \acp{ap} and corresponding anchors, geometry of the room, device orientation, etc~\cite{comst2023anish}. In practice, distributing and maintaining this information is not always feasible. 
Machine learning and deep learning techniques have also been considered for high-accuracy indoor localization~\cite{deepL2020mitsubishi}. However, they rely on collecting large training datasets, which is often burdensome, and the resulting models are often computationally complex for resource-constrained devices. 
%The authors of~\cite{anish2022wcnc} proposed a lightweight shallow \ac{nn} model to map the \ac{adoa} measurements (\acp{aoa} of \acp{mpc} computed with respect to a reference \ac{mpc}) to client location coordinates. They relieve the burdensome training data collection by exploiting the location labels obtained from a bootstrapping localization algorithm (here, we refer bootstrapping as the data annotation technique that automatically obtains labels for the training data). %This way, learn the behaviour of the bootstrapping algorithm and then estimating the location of the client at a much lower computational complexity.
%However, a single model can merely generalize to any generic environment as they only map the features specific to the environment they are trained in. 
In our previous conference paper~\cite{anish2022wcnc}, we proposed a lightweight shallow \ac{nn} model to map \ac{adoa} measurements to client location coordinates. These shallow \acp{nn} relieve the training data collection process by exploiting the location labels obtained from a bootstrapping localization algorithm (here, we refer to bootstrapping as the data annotation technique that automatically labels the training data). 
However, large environments or indoor spaces with irregular geometries require bigger models, which end up performing well only in those spaces for which they have been trained. Hence, a single model can rarely generalize to generic environments.

In this paper, we advocate that using multiple tiny \ac{nn} models to cover different sections (enclosed overlapping/non-overlapping sub-area) of an indoor space is better than training a single model. These \acp{nn}, one for a given section of the indoor space, are trained using location labels obtained from a geometric bootstrapping localization algorithm. Training multiple \acp{nn} offers two key advantages: ($i$) the models require less training data, thereby reducing both the training complexity of the \ac{nn} and that of the bootstrapping localization algorithm; ($ii$) the resulting localization scheme is device-centric and can be easily scaled up to multiple clients. 
In this work, we obtain the training labels in a self-supervised manner by resorting to JADE~\cite{palacios2017jade} as the bootstrapping algorithm~\cite{anish2022wcnc}. This is because JADE requires zero knowledge of the indoor environment, and employs \ac{adoa} measurements as input (same as our tiny \acp{nn}).  
In order to select the best \ac{nn} to localize the mobile client, we propose two schemes: one based on a \ac{kf}, that exploits the track information of the client, and another based on out-of-distribution detection (ODD), that exploits the statistical distribution of the training labels to select the best \ac{nn} for localization.

\begin{figure*}[t]
    \centering
    \includegraphics[width=1\textwidth, trim={1cm 0 1.7cm 0},clip]{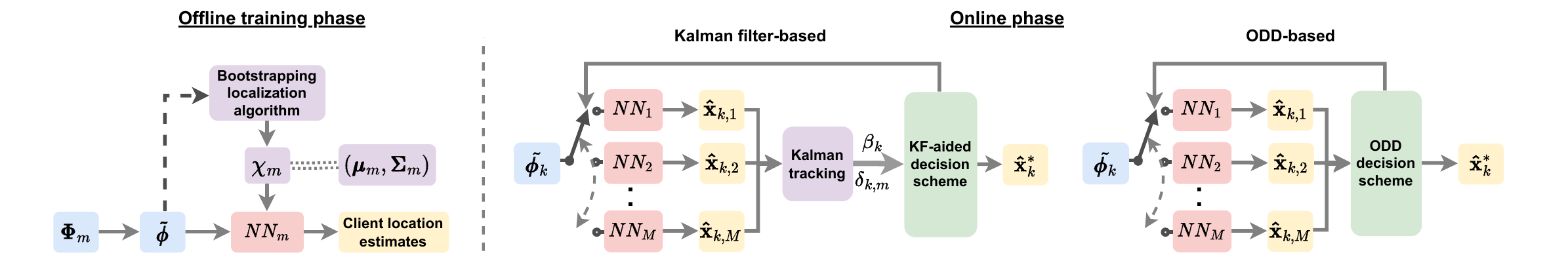}
    \caption{The workflow of our proposed approaches to select the best NN model for accurate indoor localization.}
    \label{fig:workflow}
\end{figure*}

%     Specifically, the key contributions of our work are:
% \begin{itemize} 
%     \item We propose to train multiple tiny \acp{nn} in a self-supervised fashion, by exploiting the training labels obtained from a bootstrapping localization algorithm from the literature. Each of these \ac{nn} models are trained in a specific section of a complex indoor space (here, a large and irregular shaped indoor environment).

%     \item We propose a Kalman filter tracking-based scheme to track the trajectory of the estimated client location and based on the state prediction, select the best \ac{nn} to infer the client location.
%     \item We also propose an out-of-distribution detection (ODD)-based scheme that exploits statistical distribution of training labels to switch to the right \ac{nn} for localization.

%     \item We carry out a simulation campaign to assess the performance of our proposed localization scheme and also compare it with another switching scheme that exploits the statistical distribution of the location labels (obtained from the bootstrapping algorithm) to select the best \ac{nn} for localizing the client.
    
% \end{itemize}

The outline of the paper is as follows: Section~\ref{sec:literature} presents a brief review of the existing localization schemes; Section~\ref{sec:scheme} elaborates on our proposed localization schemes; Section~\ref{sec:simulation} presents the results of our simulation campaign; finally, we draw our conclusions in Section~\ref{sec:conclusion}.

\section{Related work}
\label{sec:literature}

Geometry-based schemes exploit the angle and range information to localize \ac{mmw} devices~\cite{indoorLocSurvey2019zafari}, e.g., \acp{adoa} are used to triangulate a client's location in~\cite{palacios2019single}. %In~\cite{palacios2017jade}, the authors proposed JADE, a scheme that employs \ac{adoa} measurements to simultaneously localize the client device and the \acp{ap} with zero-knowledge about the indoor environment. However, JADE needs multiple iterations to converge to accurate location estimates, thereby increasing its computational complexity and making it infeasible to implement on resource-constrained devices. 
Blanco \emph{et al.} exploit \ac{aoa} and \ac{tof} measurements from 60-GHz 802.11ad-based and sub-6~GHz routers to trilaterate the client location~\cite{blanco2022mobisys}. The authors of \cite{polar2020Infocom} process the \ac{cir} measurements from an \ac{fpga}-based 802.11ad implementation in order to estimate \ac{tof} and \acp{aod}, and compute the client locations. They then apply a Kalman filter to smooth out the estimated trajectory.  

Deep learning methods have also been explored to localize a \ac{mmw} device. For example, \ac{rssi} and \ac{snr} beam indices from IEEE~802.11ad-based \ac{mmw} routers help learn a multi-task model for location and orientation estimation~\cite{deepL2020mitsubishi},~\cite{fing2019mitsubishi2}. The authors of~\cite{wang2023icassp} designed a dual-decoder neural dynamic learning framework to reconstruct the intermittent beam-training measurements from these routers sequentially, and thus estimate a client's trajectory. Vashisht \emph{et al.} used \ac{snr} fingerprints to train a multilayer perceptron regression model and filter these coarse estimates using a Kalman filter~\cite{kfloc}. 

\section{Proposed localization scheme}
\label{sec:scheme}

\subsection{Problem statement and main idea}

The main objective of this work is to employ multiple self-supervised tiny \acp{nn} to localize a client moving in a large indoor environment in a distributed fashion. %This is because any NN model will only be able to learn the features specific to an indoor space, especially if the features involve geometric information. 
%As observed previously~\cite{anish2022wcnc}, \acp{nn} that learn the behaviour of the bootstrapping algorithm can only mimic its performance in space they are trained in. In order to make localization using such \acp{nn} more robust and be able to localize a client in a distributed fashion. 
A key aspect then is to decide when to switch to the right \ac{nn} model, given that prior information about the ground truth locations and the map of the indoor environment is not available. 

Fig.~\ref{fig:workflow} illustrates the workflow of our proposed localization scheme, which consists of two phases. In the offline training phase, \ac{nn} $m$ (corresponding to section $m$ of a large indoor space) is trained using the \ac{adoa} values $\Tilde{\pmb{ \phi}}$ computed from the \ac{aoa} measurements ${\bf \Phi}_m$ (obtained by processing \ac{csi} at the client as in~\cite{blanco2022mobisys, mdTrack2019}). These are fed to the bootstrapping localization algorithm to obtain training labels $\chi_m$ and their corresponding mean-covariance matrix pair ($\pmb{\mu}_m,{\bf \Sigma}_m)$. These labels are used to train \ac{nn} $m$. Note that the input to both our \ac{nn} model and the bootstrapping localization algorithm are the same, i.e, $\Tilde{\pmb{\phi}}$.

For the online training phase, we propose two schemes.
The first scheme exploits the location estimates from the \ac{nn} $m$ to track the evolution of the client's state. We exploit the state innovation ${\bf y}_t$ and the \ac{nis} metric $\beta$ of the predicted state, which measures how accurately the Kalman filter predicted the measurements, to choose the best \ac{nn}.
The second scheme exploits the statistical parameters of the training labels to compute the distance between the \ac{nn} estimates from $m$ different training label distributions. The idea is to compute the distance between the \acp{nn}' estimates and the distribution of the labels (in this work, we resort to the Mahalanobis distance). As the wrongly estimated location will be far from the distribution corresponding to the true label, we refer to this scheme as \emph{out-of-distribution detection}.   

\subsection{Kalman filter (KF)-based decision scheme}
The \ac{kf}-based scheme involves two stages: the trajectory tracking phase and the decision phase.

\smallsection{Kalman tracking phase}. In this phase, we track the evolution of the client trajectory, as estimated by our trained \acp{nn}, using a Kalman filter~\cite{kalman1960new}. Let the state of the client at time $t$ be $\mathbf{s}_t  = [x_t, \dot{x}_t, y_t, \dot{y}_t]^T$, where $x$ and $y$ are the 2-D coordinates of the client, and  $\dot{x}$ and $\dot{y}$ are the $x$ and $y$ component of it's velocity. The state $\mathbf{s}$ evolves in time following a constant-velocity (CV) model. 
The evolution of the state $\mathbf{s}$ at $t$ is given as $\mathbf{s}_t = \mathbf{F}_t \mathbf{s}_{t-1} + \mathbf{w}_t$, where $\mathbf{F}_t = \pmb{I}_2 \, \scriptsize{\otimes} \left[\begin{array}{cc}
     1 & \Delta t  \\
     0& 1
\end{array}\right]$ is the state transition matrix that transforms the state of the client at time step $t-1$ to $t$, and $\mathbf{w}_t \sim \mathcal{N}(0,\mathbf{Q}_t)$ represents the zero-mean Gaussian distributed process noise with covariance matrix $\mathbf{Q}_t$. Here, $\Delta t = 1$~s, $\otimes$ is the Kronecker product, and $\pmb{I}_2$ is the $2 \times 2$ identity matrix. The predicted state representing the 2D location of the client is $\mathbf{H} \mathbf{s}_t + \mathbf{r}_t$, where $\mathbf{H}$ is the observation matrix given by diag$(1,0,1,0)$ and $\mathbf{r}_t \sim \mathcal{N}(0,\mathbf{V}_t)$ is the zero mean Gaussian observation noise, with the observation noise covariance matrix $\mathbf{V}_t$. 

The Kalman filter performs two steps: \emph{prediction} and \emph{model update}. The \emph{prediction} step estimates the current \emph{a priori} state of the client $\mathbf{s}_{t|t-1}$ based on the previous \emph{a posteriori} estimate $\mathbf{s}_{t-1|t-1}$, i.e., $\mathbf{s}_{t|t-1} = \mathbf{F}_t \mathbf{s}_{t-1|t-1} + \mathbf{w}_t$, and also computes the \emph{a priori} state covariance matrix $\mathbf{P}_{t|t-1} = \mathbf{F}_t \mathbf{P}_{t-1|t-1} \mathbf{F}^T_t$.
The \emph{model update} equations correct the existing state predictions and the covariance matrix using the measurement vector $\hat{\mathbf{x}}_{t}$ and the updated Kalman gain. The prediction error $\hat{\mathbf{y}}_{t}$ is the innovation of the Kalman filter and is given as
%%%%%%%%%%%%%%%%%%%%
% \begin{equation}
%     \mathbf{s}_{k|k-1} = \mathbf{F}_k \mathbf{s}_{k-1|k-1} + \mathbf{w}_k
% \end{equation}
% \begin{equation}
%     \mathbf{P}_{k|k-1} = \mathbf{F}_k \mathbf{P}_{k-1|k-1} \mathbf{F}^T_k
% \end{equation}

% \begin{equation}
% \mathbf{G}_k = \mathbf{H}_k \mathbf{P}_{k|k-1} \mathbf{H}^T_k + \mathbf{r}_k 
% \end{equation}
% \begin{equation}
% \mathbf{K}_k = \mathbf{P}_{k|k-1} \mathbf{H}_{k} \mathbf{G}^{-1}_k 
% \end{equation}

% \begin{equation}
%     \hat{\mathbf{y}}_{k} =\hat{\mathbf{x}}_{k} - \mathbf{H}\mathbf{s}_{k|k-1}
% \end{equation}
% \begin{equation}
% \mathbf{P}_{k|k} = \mathbf{P}_{k|k-1} - \mathbf{P}_{k|k-1}  \mathbf{K}_{k} \mathbf{H}_k
% \end{equation}    
% \begin{equation}
%     m^* = \argmin_m \delta_{k,m} ,
% \end{equation}
% \begin{equation}
%     \beta_k = \mathbf{y}^T_k \mathbf{G}^{-1}_k \mathbf{y}_k ,
% \end{equation}
% \begin{equation}
%     \mathbf{s}_{k|k} =\mathbf{s}_{k|k-1} + \mathbf{K}_k\hat{\mathbf{y}}_{k}
% \end{equation}
%%%%%%%%%%%%%%%%
\begin{equation}
    \hat{\mathbf{y}}_{t} =\hat{\mathbf{x}}_{t} - \mathbf{H}\mathbf{s}_{t|t-1} .
\end{equation}
This is used to correct the predicted state of the client as $\mathbf{s}_{t|t} =\mathbf{s}_{t|t-1} + \mathbf{K}_t\hat{\mathbf{y}}_{t}$, where $\mathbf{K}_t$ is the Kalman gain.

We exploit the innovation along with the innovation covariance $\mathbf{G}_t = \mathbf{H} \mathbf{P}_{t|t-1} \mathbf{H}^T + \mathbf{r}_t$, to compute 
\begin{equation}
    \beta_t = \mathbf{y}^T_t \mathbf{G}^{-1}_t \mathbf{y}_t .
\end{equation}
We also compute the Euclidean distance $\delta_{t}$ between the predicted state $\mathbf{H}\mathbf{s}_{t|t-1}$ and the current measurement $\hat{\mathbf{x}}_{t}$, given as $\delta_{t} = ||\mathbf{H}\mathbf{s}_{t|t-1} - \hat{\mathbf{x}}_t||_2$.  
We use $\hat{\mathbf{y}}_{t}$ and $\beta_t$ to implement the decision scheme to switch among the \acp{nn}. 

\smallsection{KF-based decision scheme.} The measurements used by the Kalman filter are the location estimates obtained from \ac{nn} $m$, i.e., the \ac{nn} model trained in section $m$ of the indoor space. \ac{nn} $m$ will be able to estimate the client location accurately, as long as the client is moving within region $m$. Thus, a Kalman filter will be able to predict the location of the client with a low $\delta_{t}$. At any time step $k$, as the client moves into another region, \ac{nn} $m$ will be unable to estimate the location of the client accurately. This is because the client will not be able to listen to all the multipath components from the \acp{ap} of the previous region. However, the Kalman filter would predict the location estimate based on the model learned up to time step $k-1$, thus the Kalman predicted state would be closer to the location of the client. As the current \ac{nn} will be unable to accurately localize the client, this would result in a large $\delta_{k}$ and an even larger $\beta_k$. When $\beta_k$ exceeds a user-defined threshold $\eta$, we feed the corresponding \ac{adoa} measurements $\Tilde{\pmb{\phi}}_k$ to the $M$ trained \ac{nn} models to compute the location estimate $\hat{\mathbf{x}}_{k,m}$, where $m = 1,\cdots,M$. We select the NN $m^*$ that minimizes the Euclidean distance between $\hat{\mathbf{x}}_{k,m}$ and the Kalman-predicted location $\mathbf{H}_k\mathbf{s}_{k|k-1}$: \begin{equation}
    m^* = \argmin_m \delta_{k,m} ,
\end{equation}
where $\delta_{k,m} = ||\mathbf{H}_k\mathbf{s}_{k|k-1} - \hat{\mathbf{x}}_{k,m}||_2$, with the best estimate of the client being $\hat{\mathbf{x}}^*_{k} = \hat{\mathbf{x}}_{k,m^*}$. The subsequent set of measurements for the Kalman filter would be the locations estimated by NN $m^*$. We repeat this procedure to switch to the right \ac{nn} whenever the $\beta$ metric exceeds a threshold $\eta$.  
In our evaluation, we set $\eta = 2$. This aligns with the theoretical result that the expectation of $\beta$ should equal the number of degrees of freedom of the Kalman filter (i.e., the 2 coordinates of the client in our case)~\cite{reid2001estimation}.

\subsection{Out-of-distribution detection (ODD) switching scheme}

In this approach, we exploit the statistical properties of the location labels used to train the \acp{nn} of each section of the indoor space. Specifically, we compute the mean and the covariance of the training labels distribution $\chi_m$ corresponding to region $m$ obtained from the offline training phase. Each distribution $\chi_m$ is characterized by its mean-covariance pair $(\pmb{\mu}_m, \pmb{\Sigma}_m)$. %, where $m$ represents a section of the indoor space.
In the online testing phase, at any time instance $k$, if the distance between the current location estimate $\hat{\bf x}_{k}$ and the previous estimate $\hat{\bf x}_{k-1}$ exceeds a user-defined threshold $\zeta$, i.e.,  $||\hat{\bf x}_{k} - \hat{\bf x}_{k-1}||_2 > \zeta$, where $\zeta$ is the distance threshold (1~m in our case), the \ac{adoa} values $\Tilde{\pmb{\phi}}_k$ at test location $k$ are fed to $M$ \acp{nn}. This results in $M$ location estimates $\hat{\bf x}_{k,m}$. We then compute the Mahalanobis distance $\rho_{k,m}$ of $\hat{\bf x}_{k,m}$ from each of the distributions $\chi_m$ as    
\begin{equation}
   {\rho}_{k,m}  = \sqrt{(\hat{\bf{x}}_{k,m} - \pmb {\mu}_m)^T \pmb{\Sigma}^{-1}_m ({\bf\hat{x}}_{k,m} - \pmb{\mu}_m)} \,.
\end{equation}
The Mahalanobis distance measures the distance of a point from a distribution of points, where, the smaller the distance, the closer the point to the distribution. %However, there are possibilities of the estimated location to be wrongly estimated close to another distribution, which may affect the choice of NN.
We finally choose the best \ac{nn} as $m^* = \argmin_m \,\rho_{k,m}$, and its corresponding estimate $\hat{\bf x}_{k,m^{*}}$ as the best location estimate. Whenever the location estimated by \ac{nn} $m^*$ is far from the previous estimate, i.e., the Euclidean distance between the current estimate and the previous estimate exceeds the distance threshold $\zeta$ (chosen after exhaustive search), we repeat the above procedure and switch to another \ac{nn} model.

\subsection{Tiny neural network architecture}
We resort to a 4-layer tiny \ac{nn} model with $(N_i, \,N_{h_1}, \,N_{h_2},\, N_{h_3},\, 2)$ neurons in each layer. Here, $N_i$ is greater than or equal to the number of \acp{adoa} from visible anchors~\cite{anish2022wcnc}, $N_{h_1} = \lceil{\kappa N_i  \rceil}$, $N_{h_2} = N_{h_1}$, $N_{h_3}=\lceil{N_{h_2}/2\rceil}$,  and $\lceil \cdot \rceil$ represents the ceiling function. The \ac{nn} has two output neurons, corresponding to the 2D coordinates of the client. The \ac{nn} learns a non-linear regression function $\mathcal{F}(\Tilde{\pmb{\phi}}_t)$ between the \acp{adoa} $\Tilde{\pmb{\phi}}_t$ and the client location $\hat{\bf x}_t$. 
The regression problem is framed to minimize the \ac{mse} between the self-supervised training labels and the locations estimates obtained from the \ac{nn}. The \ac{nn} employs the \ac{relu} activation function and Adam optimizer. The hyperparameters used for tuning the \ac{nn} to minimize the loss function are the factor $\kappa$, dropout rate $p$, learning rate $r$, and the training batch size $b$. 

\section{Simulation Results}
\label{sec:simulation}

\subsection{Simulation environment}

\begin{figure}[t]
    \centering
    \includegraphics[width=0.9\columnwidth]{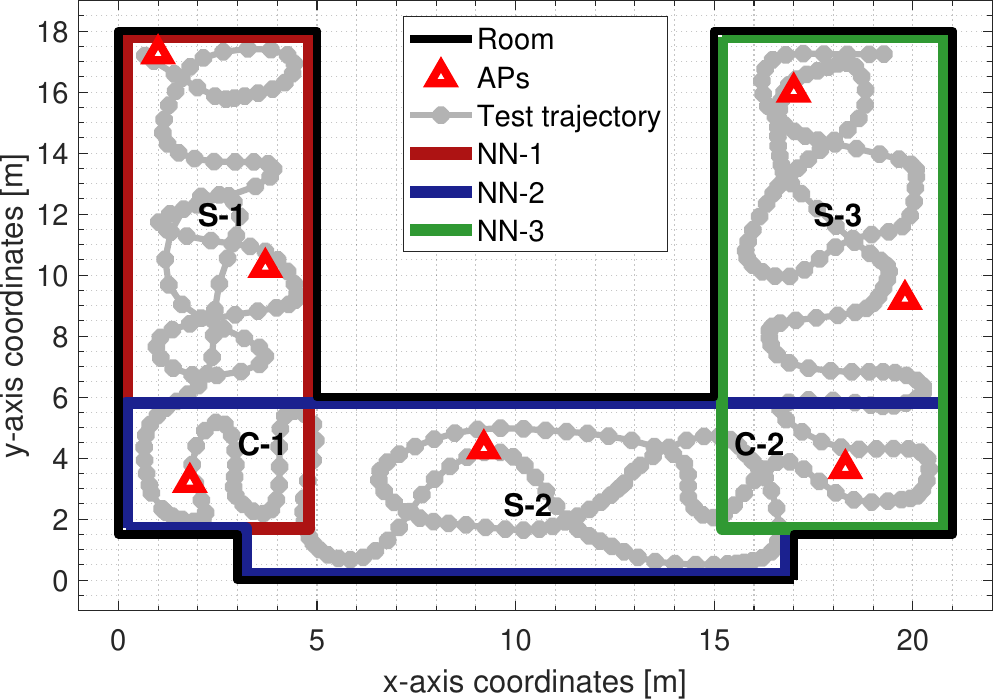}
    \caption{U-shaped room used for simulations. The colored boxes represent the three sections of the room and the grey line represents the test trajectory.}
    \label{fig:Uroom}
\end{figure}

We validate the performance of our proposed schemes via a simulation campaign. We simulate human motion trajectories in an irregular U-shaped room as shown in Fig.~\ref{fig:Uroom}, comprising two vertical rectangular sections (S-1 and S-3) of size 5~m $\times$ 18~m and 6~m $\times$ 18~m, connected by a 20~m wide horizontal section (S-2). We deploy seven \ac{mmw} \acp{ap} to cover the indoor space. %at $(1,12)$, $(2,16)$, $(9,14)$, $(8,7.5)$, $(6.5,1.5)$, $(13,4)$ and $(19,2.5)$.
We collect \acp{aoa} from all the \acp{ap} and their corresponding \acp{va} (mirror images of the physical \acp{ap} with respect to each wall of the room) at each client location along a trajectory using a ray tracer. To simulate realistic noisy measurements, we perturb the \acp{aoa} with zero-mean Gaussian noise of standard deviation $\sigma = 5^\circ$. 

We train each model with $\approx800$ locations within each boxed section (S-1, S-2, S-3) in Fig.~\ref{fig:Uroom}. The trained \acp{nn} have the following architecture: NN-1 and NN-2 have $(53, 48, 48, 24, 2)$ neurons in each layer with $\kappa = 0.9$, $p = 0.1$, $r = 0.003$, $b = 50\%$, and $\kappa = 0.9$, $p = 0.05$, $r = 0.004$, $b = 50\%$ respectively, while NN-3 has $(53, 43, 43, 22, 2)$ neurons with $\kappa = 0.8$, $p = 0.1$, $r = 0.003$, $b = 50\%$.
We test the trained \acp{nn} on a trajectory comprising 388 client locations across the three sections of the room (grey line). 

\subsection{Analysis of the switching schemes}
\begin{figure}[t]
    \centering
    \includegraphics[width=1\columnwidth]{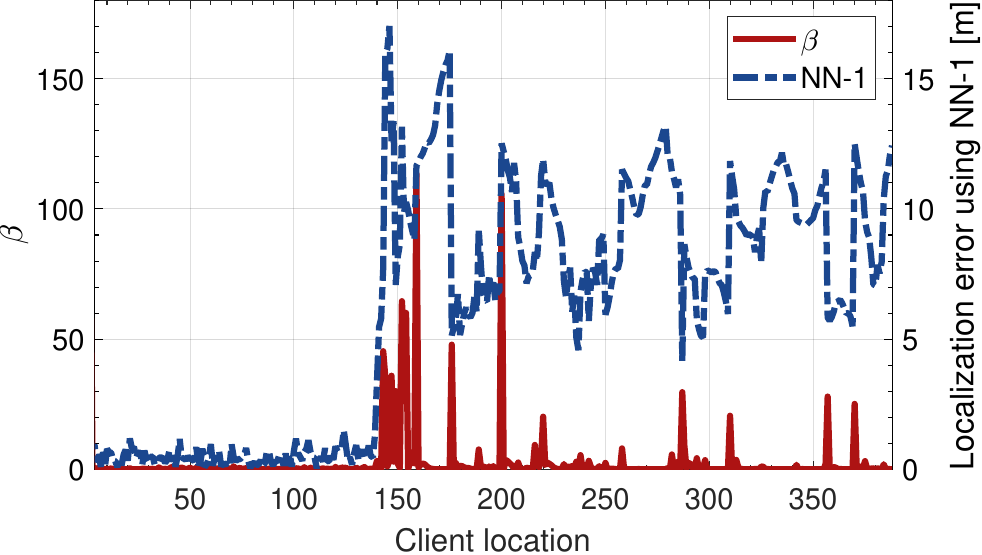}
    \caption{$\beta$ and localization error obtained when using NN-1.}
    \label{fig:beta}
\end{figure}

We first analyze the performance of the KF-based switching scheme. Fig.~\ref{fig:beta} illustrates the metric $\beta$ computed while using \ac{nn}-1 to track the client along the test trajectory. The figure also presents the localization error when using \ac{nn}-1. Initially, when \ac{nn}-1 estimates the client locations in S-1, $\beta$ is close to zero. As the client crosses the border of S-1 and S-2 (around location 140), the errors of \ac{nn}-1's estimates increase. These estimates significantly deviate from the client's state as predicted by the Kalman filter, resulting in a large $\delta_t$ and hence an even larger $\beta$ value, leading to the appearance of peaks. Conversely, with every subsequent wrong estimates by \ac{nn}-1, the Kalman filter keeps predicting the client's location to be around the wrong estimates, resulting again in low values of $\beta$. This trend can be observed when the client moves in S-2 and S-3. % Moreover, once the Kalman filter has learned the model around the wrongly estimated values, $\beta$ keeps decreasing, since the Kalman filter expects the client to  feeds on \ac{nn}-1's estimates, thus predicting correctly around the wrong location estimates. 

\begin{figure}[t]
    \centering
    \includegraphics[width=0.95\columnwidth,trim={0 0 0 4mm},clip]{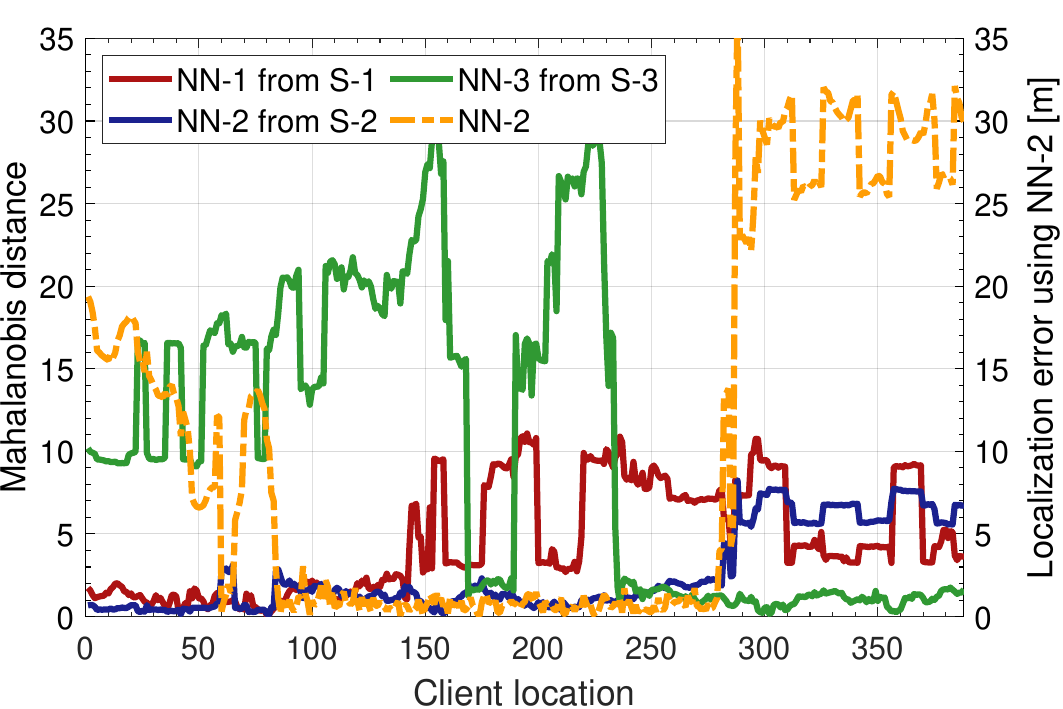}
    \caption{Mahalanobis distance between the estimated client locations and $\chi_m$ for each \ac{nn}. The dashed line represents the localization error of the test trajectory using NN-2.}
    \label{fig:mahaDist}
\end{figure}

We now analyze the ODD scheme. Fig.~\ref{fig:mahaDist} illustrates the Mahalanobis distance between the location estimates obtained from different \acp{nn} and the distribution of the training data $\chi_m$ corresponding to each section of the room, as a function of the client location. The locations estimated by the \acp{nn} and lying within the distribution of the training data consistently yield $\rho \leq 2$. We also observe that a large set of locations in the common sections of the room, i.e., C-1 (client locations from 90 to 140) and C-2 (from 235 to 270), were localized accurately by the \acp{nn} sharing the overlapped regions, as they compute the \acp{adoa} from the same set of visible anchors. Thus, these estimates have similar $\rho$ values, as they are part of two $\chi$s. In such situations, we can exploit the Euclidean distance between the current location estimate and the previous location estimate to choose the right \ac{nn}.  

While the values of $\rho$ for NN-1 and NN-3 are low in their respective regions and easily distinguishable thanks to \ac{adoa} measurements from a disjoint set of \acp{ap}, the $\rho$ values for location estimates from \ac{nn}-2 are low for the entire trajectory. This is because a vast majority of the locations in S-2 can compute the \ac{adoa} values from \acp{mpc} arriving from all the \acp{ap}. Thus, \ac{nn}-2 estimates the client locations closer to all the three $\chi$s. However, the localization error (yellow dashed line) is still large, especially for the locations in the non-overlapping regions of S-1 and S-3. This implies that NN-2's location estimates, even though highly erroneous, lie within the other two distributions. Thus, we remark that the ODD technique is more appropriate in environments where each section is illuminated by a disjoint set of \acp{ap}.

\begin{figure}[t]
    \centering
    \includegraphics[width=0.95\columnwidth]{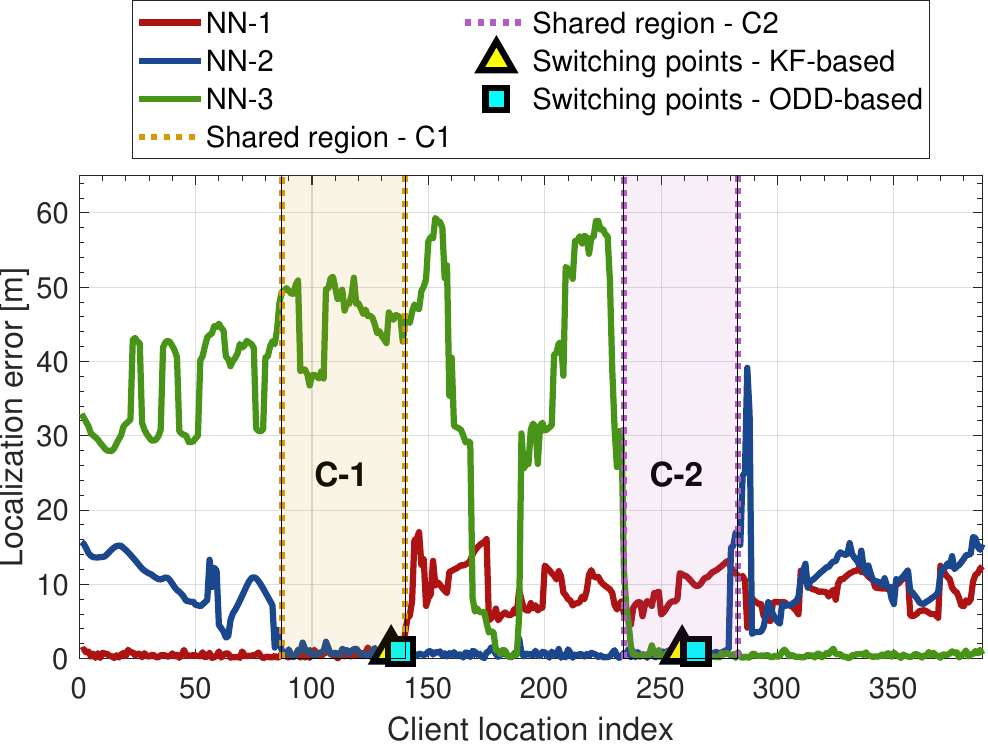}
    \caption{Localization error using the three \acp{nn} and estimated switching points using both the \ac{kf}-based (triangles) and ODD-based (squares) switching schemes. The shaded sections represent the overlap regions C-1 and C-2.}
    \label{fig:allPlot}
\end{figure}

The switching points obtained from the \ac{kf}-based (triangles) and ODD-based (squares) switching schemes are depicted in Fig.~\ref{fig:allPlot}. The dotted and shaded sections are the borders and the common areas C-1 and C-2, between the S-1 and S-2, and S-2 and S-3 respectively. The ideal switching points are within the boundaries of C-1 and C-2 (see Fig.~\ref{fig:Uroom}), from location index 90 to 140 (C-1) and from 235 to 270 (C-2). We observe that both schemes decide to switch \ac{nn} models within the ideal switching area. However, unlike the ODD scheme, the \ac{kf}-based scheme tends to be more robust, as it uses the motion model of the client to predict its future location. %This helps choosing the best location estimate among the set of location estimates from the \acp{nn} easier.

\begin{figure}[t]
    \centering
    \includegraphics[width=0.98\columnwidth]{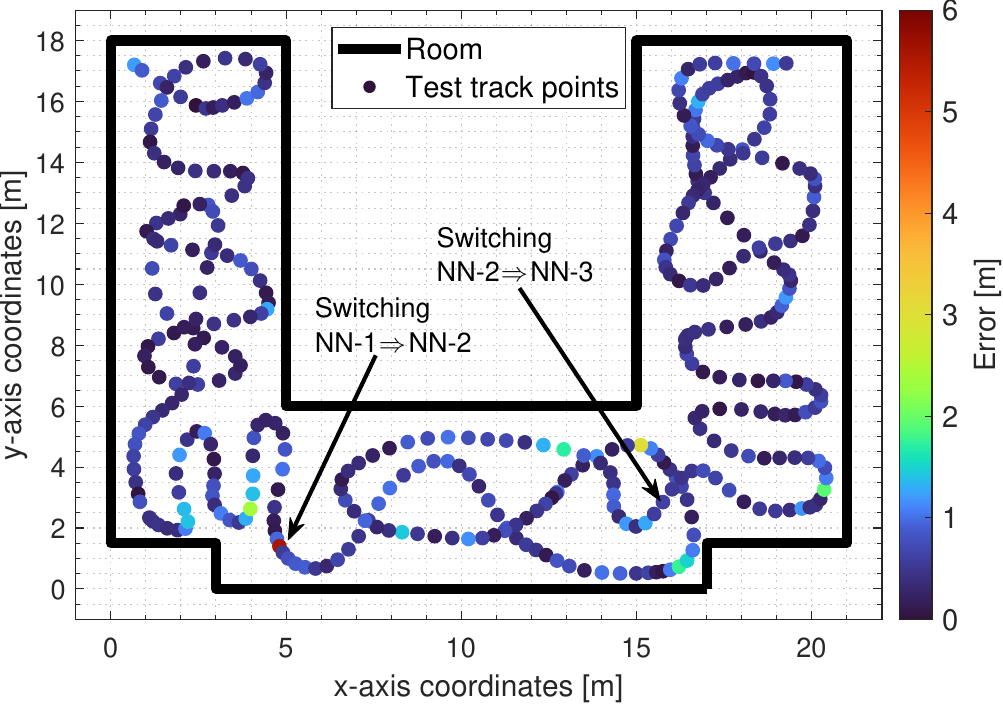}
    \caption{Location-wise estimation error after reconstructing the trajectory upon switching among the 3 \acp{nn} using the KF-based switching scheme.}
    \label{fig:track}
\end{figure}

\begin{figure}[t]
    \centering
    \includegraphics[width=\columnwidth]{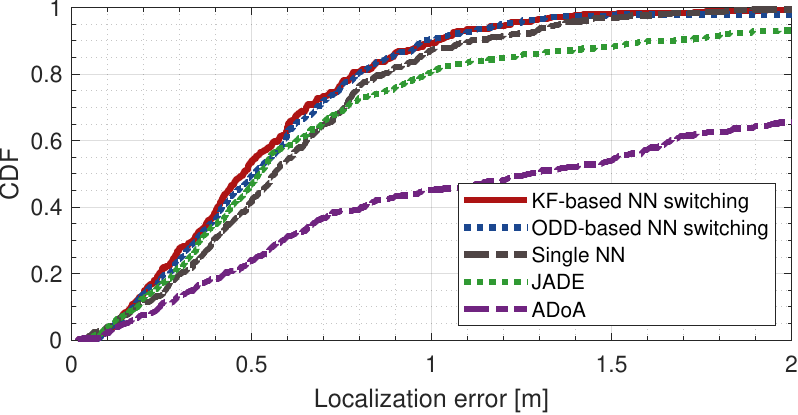}
    \caption{\ac{cdf} of the trajectory reconstructed using KF-based switching scheme against JADE.}
    \label{fig:cdf}
\end{figure}

Fig.~\ref{fig:track} illustrates the localization errors at each location along the test track, where we consider the \ac{nn} switching points obtained from the KF-based scheme, and concatenate the location estimates of each \ac{nn}. Here, blue hues represent errors $\leq 1$~m whereas green to red hues correspond to higher errors.
%, and green trajectories correspond to the location estimates obtained from NN-1, NN-2, and NN-3, respectively. Comparing with Fig.~\ref{fig:Uroom}, 
We observe sub-meter errors at most locations, and slightly larger errors near the bottom left and right corners, and at the switching points, where the \ac{nn} being used cannot accurately localize the client. %that the error at each the \acp{nn} are switched close to the edge of the boxed sections. 
Fig.~\ref{fig:cdf} presents the \ac{cdf} of the localization errors of the reconstructed trajectory and of the trajectory estimated using the JADE algorithm, a single self-supervised \ac{nn}, and the geometric \ac{adoa} algorithm~\cite{palacios2019single}. 
We observe that JADE and our self-supervised \acp{nn} outperform the \ac{adoa} scheme~\cite{anish2022wcnc}. While JADE achieves sub-meter localization errors in 80\% of the cases (mean error of 0.74~m), a single \ac{nn} performs better with sub-meter errors in about 87\% of the cases (mean error of 0.63~m). Likewise, choosing the right \ac{nn} via the \ac{kf}- or ODD-based schemes achieves sub-meter errors in about 90\% of the cases. We note that running JADE on the entire trajectory results in large errors, especially when moving closer to the bottom left and right corners of the room. This is because noisy \ac{adoa} values largely offset the estimates of the associated anchors.
%, resulting in erroneous client localization. %This is a shortcoming when training a single \ac{nn} for a large and irregular-shaped room in a self-supervised fashion.
Moreover, using a single \ac{nn} instead of multiple \acp{nn} results in large training and bootstrapping algorithm complexity, and a single \ac{nn} typically needs additional hidden layers to accurately localize a client in complex-shaped environments.
Thus, training a different, smaller \acp{nn} (each for a different portion of the environment) and employing our proposed switching schemes can help accurately localize a client in large or irregular rooms.
 
% needed in second column of first page if using \IEEEpubid
%\IEEEpubidadjcol

\section{Conclusions}
\label{sec:conclusion}
In this letter, we proposed to employ multiple self-supervised tiny \ac{nn} models to accurately localize a client as it moves across a large indoor environment. We presented two schemes based on innovation of Kalman filters and statistical distribution of training labels, to dynamically choose the best \ac{nn} model to accurately localize a client. Results show that the trajectory reconstructed upon switching to the right \acp{nn} using the proposed schemes achieves $\leq$1~m localization error in 90\% of the cases. We can conclude that both schemes can aid to dynamically switch to the best \ac{nn}. However, the KF-based scheme is more robust, owing to its ability to track client's motion, thus predicting its trajectory using \ac{nn} estimates. In contrast, the ODD scheme could be a feasible choice in large distributed environments, where different sections of the indoor space are illuminated by a disjoint set of \acp{ap}.

%\section*{Acknowledgment}

% Can use something like this to put references on a page
% by themselves when using endfloat and the captionsoff option.
\ifCLASSOPTIONcaptionsoff
  \newpage
\fi

\bibliographystyle{IEEEtran}
\bibliography{IEEEabrv,refLOC}

\end{document}